\documentclass[epj,nopacs]{svjour}
\usepackage{cite,color}
\usepackage[dvipdfm,bookmarks=false]{hyperref}
\hypersetup{colorlinks=false}
\usepackage{epsfig}
\newcommand{\nc}{\newcommand}
\nc{\am}[1]{\ensuremath{#1}}
\nc{\be}{\begin{equation}}
\nc{\ee}{\end{equation}}
\nc{\ChPT}{$\chi$PT}
\nc{\DAF}{DA\char8NE}
\nc{\GeV}{\mbox{GeV}}
\nc{\ps}{\mbox{ps}}
\nc{\mrad}{\mbox{mrad}}
\nc{\ie}{i.e.}
\nc{\MeV}{\mbox{MeV}}
\nc{\mbo}{\mathversion{bold}}
\def\cl{\centerline} \def\f{$\phi$}  \def\ab{$\sim$}  \def\dif{\hbox{d}}

\let\cl=\centerline  \def\DAF{DA\char8NE}
\def\figbox#1;#2;{\parbox{#2cm}{%
\vglue3mm\epsfig{file=#1.eps,width=#2cm}\vglue3mm}}
\def\figboxc#1;#2;{\cl{\figbox #1;#2;}}
\def\ifm#1{\relax\ifmmode#1\else$#1$\fi}  \def\f{\ifm{\phi}}   \def\epm{\ifm{e^+e^-}}  \def\ks{\ifm{K_S}}
  \def\to{\ifm{\rightarrow}}  \def\pic{\ifm{\pi^+\pi^-}}  \def\kl{\ifm{K_L}}
\def\kspp{\ks\to\pic}
\def\up#1{\ifm{^{#1}}}  \def\dn#1{\ifm{_{#1}}}  \def\ab{\ifm{\sim}}  \def\x{\ifm{\times}}  \def\minus{\ifm{-}}  \def\ts{\ifm{\tau(\ks)}}
\def\plm{\ifm{\pm}}    \def\dif{\hbox{d}}
\def\tt{\ifm{\widetilde{\tau}}}
   \def\ff{$\phi$--factory}  \def\deg{\ifm{^\circ}}  \def\bye{\end{document}}  \def\ko{\ifm{K^0}}  \def\ord#1;{\ifm{\cal O}(#1)}
\catcode`@=11
\newdimen\z@ \z@=0pt
\newskip\z@skip \z@skip=0pt plus0pt minus0pt
\def\m@th{\mathsurround=\z@}
\def\ialign{\everycr{}\tabskip\z@skip\halign}
\def\eqalign#1{\null\,\vcenter{\openup\jot\m@th
  \ialign{\strut\hfil$\displaystyle{##}$&$\displaystyle{{}##}$\hfil
      \crcr#1\crcr}}\,}
\catcode`@=12
\renewcommand\Gamma{\char0}
\renewcommand\Delta{\char1}
\begin{document}
\title{\mbo Precision Measurement of \ks\ Meson Lifetime with the KLOE detector}
\subtitle{The KLOE Collaboration}
\author{
F.~Ambrosino\inst{3,4} \and
A.~Antonelli\inst{1} \and
M.~Antonelli\inst{1} \and
F.~Archilli\inst{8,9} \and
G.~Bencivenni\inst{1} \and
C.~Bini\inst{6,7} \and
C.~Bloise\inst{1} \and
S.~Bocchetta\inst{10,11} \and
F.~Bossi\inst{1} \and
P.~Branchini\inst{11} \and
G.~Capon\inst{1} \and
T.~Capussela\inst{1} \and
F.~Ceradini\inst{10,11} \and
P.~Ciambrone\inst{1} \and
A.~De~Angelis\inst{14} \and
E.~De~Lucia\inst{1} \and
M.~De~Maria\inst{15}
A.~De~Santis\inst{6,7} \and
P.~De~Simone\inst{1} \and
G.~De~Zorzi\inst{6,7} \and
A.~Denig\inst{2} \and
A.~Di~Domenico\inst{6,7} \and
C.~Di~Donato\inst{4} \and
B.~Di~Micco\inst{10,11} \and
M.~Dreucci\inst{1} \and
G.~Felici\inst{1} \and
S.~Fiore\inst{6,7} \and
P.~Franzini\inst{6,7} \and
C.~Gatti\inst{1} \and
P.~Gauzzi\inst{6,7} \and
S.~Giovannella\inst{1} \and
E.~Graziani\inst{11} \and
M.~Jacewicz\inst{1}\and
V.~Kulikov\inst{13} \and
J.~Lee-Franzini\inst{1,12} \and
M.~Martini\inst{1,5,16} \and
P.~Massarotti\inst{3,4} \and
S.~Meola\inst{3,4} \and
S.~Miscetti\inst{1} \and
M.~Moulson\inst{1} \and
S.~M\"uller\inst{2} \and
F.~Murtas\inst{1} \and
M.~Napolitano\inst{3,4} \and
F.~Nguyen\inst{10,11} \and
M.~Palutan\inst{1} \and
A.~Passeri\inst{11} \and
V.~Patera\inst{1,5} \and
P.~Santangelo\inst{1} \and
B.~Sciascia\inst{1} \and
A.~Sibidanov\inst{1} \and
T.~Spadaro\inst{1} \and
C.~Taccini\inst{10,11} \and
L.~Tortora\inst{11} \and
P.~Valente\inst{7} \and
G.~Venanzoni\inst{1} \and
R.~Versaci\inst{1,5,17}
}
\institute{
Laboratori Nazionali di Frascati dell'INFN, Frascati, Italy. \and 
Institut f\"ur Kernphysik, Johannes Gutenberg - Universit\"at Mainz, Germany. \and 
Dipartimento di Scienze Fisiche dell'Universit\`a ``Federico II'', Napoli, Italy. \and 
INFN Sezione di Napoli, Napoli, Italy. \and 
Dipartimento di Energetica dell'Universit\`a ``La Sapienza'', Roma, Italy. \and 
Dipartimento di Fisica dell'Universit\`a ``La Sapienza'', Roma, Italy. \and 
INFN Sezione di Roma, Roma, Italy. \and 
Dipartimento di Fisica dell'Universit\`a ``Tor Vergata'', Roma, Italy. \and 
INFN Sezione di Roma Tor Vergata, Roma, Italy. \and 
Dipartimento di Fisica dell'Universit\`a Roma Tre, Roma, Italy. \and 
INFN Sezione di Roma Tre, Roma, Italy. \and 
Physics Department, State University of New York, Stony Brook, USA. \and 
Institute for Theoretical and Experimental Physics, Moscow, Russia. \and 
Universit\`a di Udine e IUAV Venezia, Italy. \and 
Universit\`a di Udine e LIP/IST, INFN Sezione di Trieste, Italy. \and 
Present Address: Dipartimento di Scienza e Tecnologie applicate, Universit\`a Guglielmo Marconi, Roma, Italy. \and 
Present Address: CERN, CH-1211 Geneva 23, Switzerland.
}
\authorrunning{The KLOE Collaboration}
\mail{Mario.Antonelli@lnf.infn.it,\\Marco.Dreucci@lnf.infn.it}
\date{Received: date / Revised version: date}
\abstract{
Using a large sample of pure, slow, short lived \ko\ mesons collected with KLOE detector at \DAF, we have measured the \ks\ lifetime. From a fit to the proper time distribution we find $\ts = (89.562 \pm 0.029_{\rm{stat}}\pm 0.043_{\rm{syst}})$ ps. This is the most precise measurement today in good agreement with the world average derived from previous measurements. We observe no dependence of the lifetime on the direction of the \ks.
} 
\titlerunning{Precision Measurement of \ks\ Meson Lifetime with the KLOE detector}
\maketitle
\section{Introduction}
\label{sec:intro}
We have collected very large samples, \ord10\up9 events;, of slow $K$-mesons of well known momentum, with the KLOE detector at \DAF. Kaons originate from the decay of \f-mesons produced in \epm\ collisions. We have used the above samples to measure many properties of kaons such as masses, branching ratios and lifetimes, refs. \citen{br1} through \citen{kaonmass}. The ultimate motivation was the determination of the quark mixing parameter $V_{us}$, see ref. \citen{vus}. KLOE had not however attempted to measure the \ks\ lifetime.
We pre\-sent a precise measurement of the \ks\ lifetime based on a sample of about 20 million \kspp\ decays corresponding to an \epm\ integrated luminosity of 0.4 fb$^{-1}$.

The reaction chain \epm\to\f, \f\to\kl(unobserved)\ks, \kspp, with $p_\phi$ = 13 MeV in the horizontal plane, is geometrically and kinematically overdetermined. We can therefore, event by event, determine the \ks-meson vector momentum {\bf p\dn K}, the kaon production point P and its decay point D. From $p$\dn K, P and D we obtain the decay proper time of the \ks. A fit to the proper time distribution gives the \ks-meson lifetime. The vast available statistics allows us to select some 20 million \kspp\ decays with favorable configuration to provide the most accurate and least biased measurement of time. Averaging over the sample gives a statistical accuracy of \ab2 $\mu$m in the measurement of the kaon mean decay length. For consistency we use our value of the kaon mass, $M_K$=(497.583 \plm 0.021) MeV, ref. \citen{kaonmass}.

The KLOE detector has been described in all the references mentioned above, see also refs. \citen{dc}, \citen{emc}, \citen{trg}, \citen{ncim}. In particular ref. \citen{vus} summarizes the use of the KLOE detector in collecting kaon data and reconstructing all decay channels.
\section{Data reduction}
\label{sec:ana}
Data were collected in 2004 with the KLOE detector at \DAF, the Frascati \ff. \DAF\ is an \epm\ collider operating at a center of mass energy $\sqrt s$\ab1020 MeV, the \f-meson mass. Beams collide at an angle of $\pi$-0.025 rad. For each run of about 2 hours, we measure the CM energy $\sqrt s$, ${\bf p}_\phi$ and the average position of the beams interaction point P using Bhabha scattering events. Data are combined into 34 run periods each corresponding to an integrated luminosity of about 15 pb\up{-1}. For each run set, we generate a sample of Monte Carlo (MC) events of \ab3\x\ equivalent statistics. We use a coordinate system with the $z$-axis along the bisector of the external angle of the \epm\ beams, the so called beam axis, the $y$-axis pointing upwards and the $x$-axis toward the collider center.\\
\kspp\ decays are reconstructed from two opposite sign tracks which must intersect at a point D with $r_{\rm D}\!<\,$10 cm and $|z_{\rm D}|\!<\,$20 cm, where $x\!=\!y\!=\!z\!=0$ is the \epm\ average collision point. The invariant mass of the two tracks, assumed to be pions, must satisfy $|M_{\pi\pi}-M_{K^0}|\!<\!5$ MeV. D is taken as the decay point.
The kaon momentum ${\bf p}_K$ can be obtained from the sum of the pion momenta and also from the kaon direction with respect to the known, fixed \f\ momentum ${\bf p}_\phi$. We call the latter value ${\bf p}^\prime_K$. The magnitude of the two values of the kaon momentum must agree to within 10 MeV. If the two tracks intersect in more than one point satisfying the above requirements, the one closest to the origin is retained as the \ks\ decay point. We refer to the finding of D as vertexing.

\begin{figure}[ht]
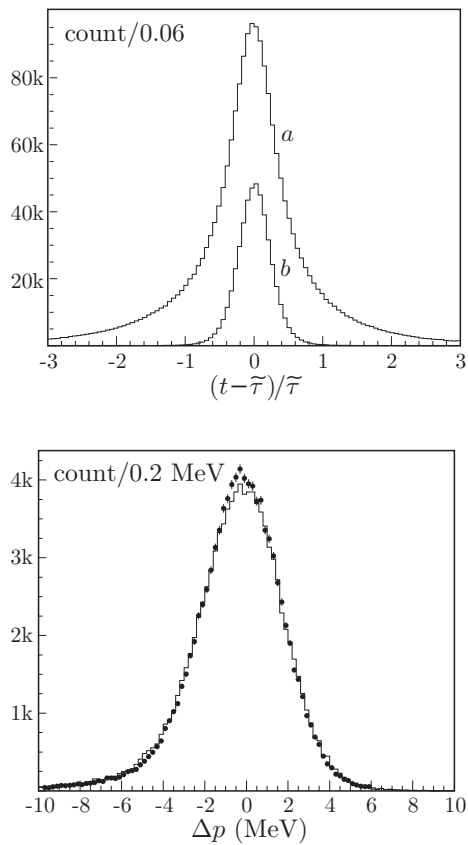

\cl{\figbox resoa;6;}
\cl{\figbox dpa;6;}
\caption {Top. Monte Carlo. Reconstructed time resolution, histogram a, in units of \tt\ for the initial \ks\ sample (rms spread \ab0.86\tt) and after cuts and geometrical fit, b, (rms spread \ab0.32\,\tt). Bottom. Distribution of $\Delta p$=$p_K-p_K^ {\,\prime}$ for data (dots), and Monte Carlo (line). The tail at left is due to the initial state radiation.}
\label{fig:reso}
\end{figure}

The above procedure selects a \kspp\ sample almost 100\% pure.
For each event we need the kaon production point P. In fact only the $z$-coordinate of P is required since the interaction region is 2-3 cm long while the other dimensions are negligible and the $x,\ y$ coordinates well known.
P lies on the beam axis and is taken as the point of closest approach to the \ks\ path as determined by the \pic\ tracks.
The resolution in $z_{\,\rm P}$ is about 2 mm. Events with $|z_{\,\rm P}|\!>\,$2 cm are rejected. From the length of PD and $p^\prime_K$ we compute the proper time in units of a reference value of \tt, the lifetime value used in our MC, \tt=89.53 ps. Its distribution is shown in fig. \ref{fig:reso} top, histogram a.
The distribution has an rms spread of 0.86 \tt\ and is not symmetric.
Time resolution can be improved discarding events with poor vertexing resolution. From MC we observe that bad vertex reconstruction is correlated with large values of $\Delta p$ = $p_K-p^{\,\prime}_K$, the difference in magnitude of ${\bf p}_K$ and ${\bf p}^\prime_K$. Fig.\,\ref{fig:reso} bottom shows the $\Delta p$ distribution for data and MC. We therefore retain events with $\cos\alpha_{\pi\pi}<-0.87$, $0.5<|\alpha^\perp_{\pi^+K}|<2.2$ rad, $|M_{\pi\pi}-M_K|\!<\,$ 2 MeV and events with $-0.5<\cos\theta(\pi^\pm)<+0.5$. $\alpha_{\pi\pi}$ is the opening angle of the pion pair. The definition of $\alpha^\perp_{\pi^+K}$ is slightly more complicated. Information about the angle between the positive pion and the kaon at the decay point D is required. We must also distinguish between the two \pic\ 'V' configurations illustrated in fig. \ref{tplg}.
\begin{figure}[htb]\cl{\figbox 2tplgy;5;}\caption{The two configurations for a \ks\to\pic\ decay.}\label{tplg}\end{figure}
Calling {\bf r} and {\bf s} the projections of kaon and positive pion on the $\{x, y\}$ plane, $\alpha^\perp_{\pi^+K}$ is defined as
$$\alpha^\perp_{\pi^+K}={\rm{sign}}\,\left(({\bf r}\times{\bf s})_z\right)\arccos\left(\frac{{\bf r}\cdot{\bf s}}{rs}\right).$$
The angle $\alpha^\perp_{\pi^+K}$ is defined in $\{-\pi,\ \pi\}$. Positive sign corresponds to the configuration of fig. \ref{tplg}, left. All  angles are in the laboratory system.

\begin{figure}[htb]
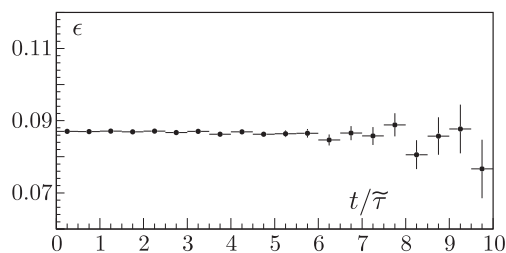
\cl{\figbox efftot;6.5;}\caption{Monte Carlo: final efficiency, averaged over all \ks\ directions, as a function of the proper time in a single run period.}\label{fig:eff}\end{figure}

After applying the cuts above, only \ab1/3 of the events survive while the rms time spread is reduced to 0.63 \tt.
Another significant improvement is obtained performing a geometrical fit of each event to obtain the production point P and the decay point D. We chose a new point P\up{\,\prime} on the beam axis and a new decay point D\up{\,\prime} on a line through P\up{\,\prime}, parallel to the kaon path, so as to minimize the $\chi^2$ function
$${|{\bf r}_{\rm D^{\,\prime}}-{\bf r}_{\rm D}|^2\over\sigma^2_{r_{\rm D}}} + {(z_{\rm P^{\,\prime}}-z_{\rm P})^2\over\sigma^2_z}.$$ The proper time distribution, after all cuts and the fit, is shown in fig. \ref{fig:reso} top,  curve b. The rms spread in $t$ is 0.32\,\tt.
We check the correctness of the \ks\ direction using a sample of \kl-mesons reaching the calorimeter, where they are detected by nuclear interactions. The \kl\ interaction point in the calorimeter together with the known $\phi$ momentum gives the \ks\ direction with good resolution. Comparison with the \ks\ direction as obtained from pions shows a negligible difference. The final efficiency for \kspp\ detection is shown in fig. \ref{fig:eff} as a function of proper time.

The average efficiency depends on the \ks\ direction, is almost flat and in average is \ab9\%.
Errors in the reconstruction of the pion tracks can bias the position of P and D. In fact, the value of \ks\ lifetime differs by \ab6\% for events with $\alpha^\perp_{\pi^+K}>0$ and $<0$, where the sign distinguishes the topologies of the di-pion `V'. see fig. \ref{tplg}.

We do correct for this effect. From MC we obtain the correction, $\Delta\ell_K$, to be applied to the \ks\ decay length, as a function of $\Delta p$. The correction is applied event by event to the data. The procedure is repeated for each run period. After applying this correction the 6\% difference mentioned above is reduced to \ab\ $10^{-3}$, although the average result is only \ab2$\sigma$ (0.1\%) different from the result before applying it.
\section{Proper time distribution fit}
\label{sec:fit}
MC and data, see fig. \ref{fig:reso} top, studies show that the time resolution is well described by the sum of two Gaussians. We write the resolution function, normalized to unity, as
$$\eqalign{&r(t,\tau,\sigma_1,\sigma_2,\alpha)=\cr
&\kern3mm{\alpha\over\sigma_1\sqrt{2\pi}}\,\exp \left(-{t^2\over2\,\sigma_1^2}\right)+ {1-\alpha\over\sigma_2\sqrt{2\pi}} \,\exp\left(-{t^2\over2\,\sigma_2^2}\right)\cr}$$
and the decay function, for a lifetime $\tau$, as:
$$d(t)={1\over\tau}\x\exp\left(-{t\over\tau}\right)\x\theta(t).$$
The expected decay curve, normalized to unity, is given by the convolution
$$g(t)=\int_{-\infty}^{\infty}d(\eta)\,r(t-\eta)\,\dif\eta.$$ Allowance must be still be made for small mistakes in the reconstruction of the decay and production position, D and P. A shift $\delta$ in the proper time is therefore introduced. Thus the function which we use for fitting the observed distribution is
$$f(t,\tau,\sigma_1,\sigma_2,\alpha,\delta)=g(t-\delta).$$
The four parameters, $\sigma_1$, $\sigma_2$, $\alpha$, $\delta$ in $f(t)$ depend on colatitude and azimuth, $\theta$ and $\phi$, of the kaon and it is not realistic to attempt to obtain them from MC. We divide the data in a 20\x18 grid in $\cos\theta,\phi$ and fit each data set for the lifetime $\tau$ with the above parameters free. In order to improve the result stability, we retain only events with $|\cos\theta|<0.5$ and 0$\,<\!\phi\!<\,$360\deg, discarding in this way only \ab8\% of the events. We therefore perform 180 independent fits only to events in a 10\x18 grid. The fit range, \minus1 to 6.5 \tt, is divided in 15 proper time bins.
The kaon lifetime is obtained as the weighted average of the 180 $\tau_i$ values
$$\tau(\ks)=\langle\tau\rangle=\sum_{i}{\tau_i\over\sigma^2(\tau_i)}\left/ \sum_{i}{1\over\sigma^2(\tau_i)}\right..$$
The corresponding $\chi^2$ value is
$\chi^2 = \sum_i(\tau_i-\langle\tau\rangle)^2/\sigma^2(\tau_i)$. We find  $\chi^2/\rm{dof}$ = 202/179 for a confidence level, CL, of 11.4\%. The normalized residuals of the 180 fit values $\tau_i$ have an rms spread of 1.1. Tab. \ref{tab:correl} gives the average correlations between fit parameters and fig. \ref{fig:fit} top shows a fit example.
\begin{table}[ht]\centering\begin{tabular}{l|cccc}
&  $\sigma_1$ & $\sigma_2$ & $\alpha$ & $\delta$ \\\hline
$\tau_S$   & 0.18 & 0.09 & 0.11 & 0.62 \\
$\sigma_1$ &      & 0.50 & 0.75 & 0.28 \\
$\sigma_2$ &      &      & 0.69 & 0.11 \\
$\alpha$   &      &      &      & 0.16 \\
\end{tabular}\caption{Correlation of fit parameters (averaged values).}
\label{tab:correl}\end{table}
\begin{figure}[ht] \cl{\figbox fit;6;}\cl{\figbox resgri;6.5;}
\caption {Top: an example of the fit with $\chi^2/\rm{dof}$= 8/10, dots are data, the line is the fit result. Bottom: proper time resolution, in units of \tt, as a function of $\phi$ and $\cos\theta$.}
\label{fig:fit}\end{figure}

The resolution ({\small{$\sqrt{\alpha\sigma_1^2+(1-\alpha)\sigma_2^2}$}}) versus $\{\theta,\phi\}$ is\break shown in fig. \ref{fig:fit} bottom. The resolution varies from 0.22\,\tt\ to \ab0.27\tt\ over the accepted $\{cos\theta,\phi\}$ range  with an average of 0.24 \tt.
The $\delta_i$ values show a dependence on $\phi$ with period $2\pi$ corresponding to a shift of the position of P of \ab$-10\,\mu$m in $y$ and \ab50$\,\mu$m in $x$. In addition, a very small, $10^{-4}$,
eccentricity of the drift chamber is evident. All these effects are consistent with mechanical and surveying inaccuracies.
To ensure that the lifetime evaluation is correct to the 10\up{-4} level, we correct the value of $\langle\tau\rangle$ obtained above by the factor $\tt/\tau_{\rm fit}^{\rm MC}$=1.00036\plm0.00019, where $\tau_{\rm fit}^{\rm MC}$ is the result of fitting the MC data with the procedure described above.
\section{Systematics and result}
\label{sec:syst}
Changes in analysis cuts and FV corresponding to a\break \ab\plm60\% change in efficiency result in a lifetime shift of 0.024\,ps. Varying the fit range gives a shift of 0.012\,ps.
As mentioned in Sec.\ref{sec:ana}, we use the KLOE value of the kaon mass in the kinematic determination of the \ks\ momentum and the calculation of $\beta_K$. The measurement of $\beta_K$ and the decay position are independent. The uncertainty on the calibration of $p^\prime_K$ gives an uncertainty of 0.033\,ps. The uncertainty due to \ks\ mass is 0.004\,ps. All fits are then performed assuming uniform efficiency versus proper time, resulting in an uncertainty of 0.005\,ps.
Table \ref{tab:syst} summarizes all systematic errors.
\begin{table}[ht]
  \begin{center}
    \begin{tabular}{lc}
      \hline
      source               & absolute value (\rm{ps})\\
      \hline
      cuts \& FV           & 0.024   \\
      fit range            & 0.012   \\
      $p^\prime_K$ calibration    & 0.033 \\
      kaon mass            & 0.004   \\
      efficiency           & 0.005   \\
      \hline
      total                & 0.043   \\
    \end{tabular}
    \caption{Systematic error contributions.}
    \label{tab:syst}
  \end{center}
\end{table}
The result is stable across the entire data taking period. As said before, without applying the vertex correction the result still remains within $2\,\sigma$ of the final result, but stability with the run period is lost. Our result for \ks\ lifetime is:
\begin{equation}
  \tau(\ks) = 89.562 \pm 0.029_{\rm{stat}} \pm 0.043_{\rm{syst}}\ \, {\rm{ps}.}
  \label{eq:life}
\end{equation}
Subdividing the data in 9 $\phi$ intervals and summing over $\cos\theta$ the $\phi$ dependence of the lifetime becomes quite obvious. The average of the 9 $\tau(\ks)$ values are of course exactly as eq. \ref{eq:life} but $\chi^2$/dof=24/8 for a CL of \ab0.2\%. Enlarging the statistical error by a factor $\sqrt{24/8}$ restores $\chi^2$=8 (CL=43\%) and corresponds to $\tau(\ks)$=89.562\plm0.050, an error very close to $\sqrt{0.029^2+0.043^2}$=0.052 confirming our estimate of the systematic error in eq. \ref{eq:life}.

\begin{figure}[ht]
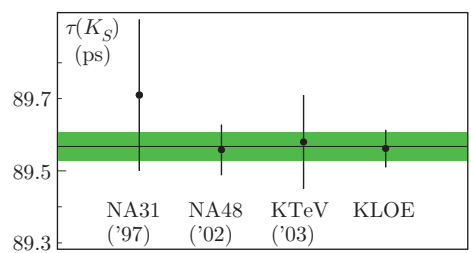
\centering\figbox wave;6.;\caption{Recent \ks\ lifetime measurements. The green band represents the new world average, $\tau(\ks)$=89.567\plm0.039 ps.}\label{fig:market}\end{figure}

The result of eq. \ref{eq:life} is in agreement with recent measurements, ref. \citen{na31,na48,ktev}, as shown in fig.\,\ref{fig:market}.
Including the present measurement, the new world average for the \ks\ lifetime is $\tau_S$=89.567\plm0.039 ps, with $\chi^2/\rm{dof}$ = 0.5/3, or CL\ab92\%.

In KLOE we can measure the lifetime for kaons traveling in different directions. We choose three orthogonal directions, the first being $\{\ell_1,\ b_1\}$ = \{264\deg, 48\deg\} in galactic coordinates. This is the direction of the dipole anisotropy of the cosmic microwave background (CMB), ref. \citen{wmap}. The other two directions are taken as $\{\ell_2,\ b_2\}$ = \{174\deg, 0\deg\} and $\{\ell_3,\ b_3\}$ = \{264\deg, -42\deg\}. After transforming the kaon momentum to the above systems, we retain only events with {\bf p}\dn K inside a cone of 30\deg\ opening angle, parallel (+) and antiparallel (\minus) to the chosen directions and evaluate the kaon lifetime. The 6 results are consistent with eq.\,\ref{eq:life}.
Defining the asymmetry ${\cal A}=(\tau^+_S - \tau^-_S)/(\tau^+_S +\tau^-_S)$, we obtain the results of tab.\,\ref{tab:cmbresult}.
\begin{table}[htb]\centering\begin{tabular}{c|c}
$\{\ell,\ b\}$              &   ${\cal A}\times10^3$ \\ \hline
\{264, 48\} &   \minus 0.2 \plm1.0\\
\{174,  0\} &          0.2\plm1.0\\
\{264,-42\} &          0.0\plm0.9\end{tabular}
\caption{Observed asymmetry. Errors are dominated by statistics.}\label{tab:cmbresult}\end{table}
Systematic errors are strongly reduced when evaluating the asymmetry. Results in tab.\,\ref{tab:cmbresult} show all the asymmetries values are well consistent with zero.

A further check has been performed using all KLOE data sample (about 2 fb$^{-1}$). The result for the asymmetry in the direction of CMB anisotropy, consistent with that given in tab.\,\ref{tab:cmbresult}, is $(\minus 0.13 \plm 0.40_{\rm{stat}})\times 10^{-3}$. No estimate of systematic error has been performed.\\

\vspace{.1cm}
{\bf Acknowledgements}.
We thank the \DAF\ team for their efforts in maintaining low background running
conditions and their collaboration during all data-taking.
We want to thank our technical staff:
G.F.~Fortugno and F.~Sborzacchi for their dedication in ensuring
efficient operation of the KLOE computing facilities;
M.~Anelli for his continuous attention to the gas system and detector safety;
A.~Balla, M.~Gatta, G.~Corradi and G.~Papalino for electronics maintenance;
M.~Santoni, G.~Paoluzzi and R.~Rosellini for general detector support;
C.~Piscitelli for his help during major maintenance periods.
This work was supported in part by EURODAPHNE, contract FMRX-CT98-0169;
by the German Federal Ministry of Education and Research (BMBF) contract 06-KA-957;
by the German Research Foundation (DFG),`Emmy Noether Programme', contracts DE839/1-4.

\end{document}